\documentstyle[prb,eqsecnum,multicol,aps,epsf]{revtex}
\begin{document}
\draft
\title{
Mesoscopic Fluctuations of Adiabatic Charge Pumping in Quantum Dots
}
\author{T.A. Shutenko}
\address{Physics Department, Princeton University, Princeton, NJ 08544}
\author{I.L. Aleiner}
\address{Department of Physics and Astronomy,
SUNY at Stony Brook, Stony Brook, NY 11794}
\author{B.L. Altshuler}
\address{Physics Department, Princeton University, Princeton, NJ 08544\\
NEC Research Institute,
4 Independence Way, Princeton, NJ 08540}
\date{Draft: \today}
\maketitle

\begin{abstract}
We consider the adiabatic charge transport through zero-dimensional
mesoscopic sample (quantum dot) caused by two periodically changing
external perturbations. Both the magnitude and the sign of the
transmitted charge are extremely sensitive to the configuration of the
dot and to the magnetic field. We find the correlation function
characterizing the random value of this pumped charge for arbitrary
strength of the perturbation. In contrast to previous theoretical and
experimental claims, the charge is found not to have any symmetry with
respect to the inversion of magnetic field. At strong pumping
perturbation, the variance of the charge $\langle Q^2\rangle$ is found 
to be proportional to the length of the contour in parametric space.
\end{abstract}


\begin{multicols}{2}
\section{Introduction}
\label{sec:1}

Adiabatic charge pumping occurs in a system subjected to a very slow
periodic perturbation. Upon the completion of the cycle, the
Hamiltonian of the system returns to its initial form; however, the
finite charge can be transmitted through a cross-section of the
system. If the Hamiltonian depends on only one
parameter, which is a strictly periodic function of time $t$, 
 the value of the charge transfer,
 $Q$ is zero. This
may be not the case for Hamiltonians, which depend
 on two or more parameters.
The pumped charge may be finite
if these parameters  follow a 
closed curve
in the parameter space,  which encompasses a finite area.
 Such an evolution may be also
characterized by a multivalued variable (angle); in this case
transmitted charge is proportional to the winding number for this
variable.
{\rm }
The value of $Q$ is not universal.
 Thouless\cite{Thouless83} showed
that for certain one dimensional systems with a gap in the
excitation spectrum in the
thermodynamic limit the charge $Q$ is quantized.
Such quantized charge pumping could be of practical importance
as a standard of electric current\cite{Niu}. The accuracy of
charge quantization depends on how adiabatic  the
process is.

The practical attempts of creating a  quantized electron pump
are based on the phenomenon of
Coulomb blockade\cite{Likharev,Devoret}.
In this kind of devices, several single
electron transistors (SET) are connected in series to increase
the accuracy of charge quantization. At least two
SET's are necessary to obtain a non-zero charge
transfer.

Semiconductor based quantum dots\cite{review}
are often used now as Coulomb blockade devices. 
 The
advantage of these devices is the possibility of changing 
independently  the
gate voltage (and thus the average electron number in the dot)
 and
the conductance of the quantum point contacts (QPC's) separating the
dot from the leads. By doing so, one can traverse from almost
classical Coulomb blockade to a completely open dot where the
effects of the charge quantization are diminished. It was shown
theoretically\cite{AleinerAndreev}, that even weak backscattering in
one-channel QPC's leads to almost quantized value of the pumped
charge in low temperature limit, $T\to 0$.

With the further opening of the QPCs the Coulomb blockade type
charging effects become negligible\cite{BrouwerAleiner}. In this case,
the main contribution to the pumped current is associated with the
quantum interference within the quantum
dot\cite{Spivak95,Zhou99,Brouwer99}.  Mechanism related to 
the inelastic processes  was first
considered in Ref.~\onlinecite{Spivak95}. 
 In open systems, this
mechanism, however, is not effective, as the inelastic
scattering time $\tau_\epsilon$ can be much
larger than the dwell time of the electron in the dot.  In this
situation, the main mechanism of pumping is not related to the
inelastic processes but rather to change in the phase factors of the
corresponding scattering matrix \cite{Zhou99,Brouwer99}.
 Predictions of
Refs.~\onlinecite{Zhou99,Brouwer99} were apparently in accord with the
recent data of Switkes {\em et. al.}\cite{Marcus99}.

However, both papers \cite{Zhou99,Brouwer99} 
are devoted to the weak pumping regime, in which the DC current
is {\em bilinear} in the pumping amplitudes. On the other hand,
 in the experiments 
\cite{Marcus99} the pumping was not weak. 
 Our
purpose here is to construct a theory of the quantum pumping
of a finite amplitude.
 The ensemble average of the pumping charge 
 $\langle Q \rangle$ is equal
to zero.
We demonstrate that the variance of the pumping charge, 
$\langle Q^2\rangle$ increases as the square
 of the area in the
parameter space as long as $\langle Q^2\rangle\ll 1$
. With further
increase of the area, $\langle Q^2\rangle$
 increases much slower (as the
length of the contour in the parametric space).
These results are obtained  in Sec.~\ref{sec:2}.

Another important problem to clarify
is the sensitivity of $Q$ to applied magnetic field. Zhou {\em
et. al.}\cite{Zhou99}, claimed the symmetry of the pumped charge,
similar to the Onsager relation for the conductance\cite{Onsager,Bonsager}.
The direct consequence of such symmetry is that the variance of 
of pumped charge is larger in the presence of time reversal symmetry,
$H=0$ than in its absence,$
\langle Q^2(B\to \infty)\rangle \leq \frac{1}{2}\langle Q^2(B=0)\rangle
$
similarly to the conductance fluctuations\cite{AltshulerKhmelnitskii}.
Both those statements are supported by the experiment of Switkes {\em et.
al.}\cite{Marcus99}.

In our opinion, the statement about such symmetry is not 
correct.  We discuss this problem in detail in
Sec.~\ref{sec:3}. By manipulations with exact ${\cal S}$ -matrix we
 show that contrary to the conductance, the pumped charge is
{\em not} symmetric  with respect to inversion of magnetic
field:  $Q(B) \neq Q(-B)$. Moreover, we show that at large values
of the magnetic field the $Q(B)$ and $Q(-B)$ are not
even correlated $\langle Q(B)Q(-B)\rangle \to 0$. The
variance of the pumped charge is found to be independent of magnetic
field in agreement with the results of Ref.~\onlinecite{Brouwer99} for
large number of channels, and in
disagreement with Refs.~\onlinecite{Zhou99,Marcus99}.

Our findings are summarized in Sec.~\ref{sec:4}. In the same section,
we will discuss existing experiment\cite{Marcus99}.

\section{Mesoscopic fluctuations of pumping in zero magnetic field}
\label{sec:2}

\subsection{General formalism.}
\label{sec:2.1}
Our analysis will be based on the general expression for the pumped
charge derived by Brouwer \cite{Brouwer99} based on the approach of
Ref.~\onlinecite{Buttiker}. Consider the  sample connected to two
leads, as in Fig.~\ref{Fig1}. Each lead is characterized by its
transverse modes $\alpha$, where 
$1\leq \alpha\leq N_l$ for the left lead and
$N_l+1\leq \alpha\leq N_{l}+N_{r}=N_{ch}$  for the right lead. 
\narrowtext
{\begin{figure}[ht]
\vspace{0.2cm}
\epsfxsize=6cm
\hspace*{0.5cm}
\centerline{\epsfbox{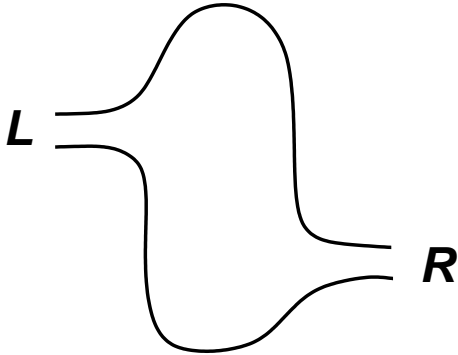}}
\vspace*{1cm}
\caption{Schematic picture of the sample connected to the leads ``L''
and ``R''.
}
\label{Fig1}
\end{figure} 
}
The sample, therefore, is completely characterized by its unitary $N_{ch}
\times N_{ch}$
scattering matrix, $\hat{\cal S}(E)$, connecting ingoing and outgoing
waves,
with energy $E$.
For instance, the two terminal conductance of the system $G$ can be
found from Landauer-Buttiker formula as
\begin{equation}
G = \frac{e^2}{2\pi\hbar}\int
dE \left(-\frac{\partial f}{\partial E}\right)
{\mathrm Tr}\left\{
\hat{\tau}_l
\hat{\cal S}(E)\hat{\tau}_r
\hat{\cal S}^\dagger (E)\right\},
\label{landauer}
\label{eq:2.1}
\end{equation}
where $f(E)=1/(1+e^{E/T})$ is the Fermi distribution function (energy
$E$ is measured from the Fermi level). Matrices $\hat{\tau}_{l(r)}$ are
projectors on the states of left (right) lead:
\begin{eqnarray}
\left[\hat{\tau}_{l}\right]_{\alpha\beta}
&=&\delta_{\alpha\beta}\times
\left\{\matrix{1, & 1 \leq \alpha \leq N_{l}\cr
0, & N_l < \alpha \leq N_{ch}
},\right.\label{eq:2.2}\\ 
\left[\hat{\tau}_{r}\right]_{\alpha\beta}&=&
\delta_{\alpha\beta} -
\left[\hat{\tau}_{l}\right]_{\alpha\beta}
. \nonumber
\end{eqnarray}

Let us now assume that ${\cal S}$ matrix is changing due to two
external parameters $X_{1,2}$ slowly varying with time.
The pumped charge $Q$  is given by\cite{Brouwer99}
\begin{mathletters}
\label{eq:2.3}
\begin{eqnarray}
Q= \frac{e}{\pi}
\int dE \left(-\frac{\partial f}{\partial E}\right)
\int_A dX_1dX_2
\Pi \left(E,\mbox{\boldmath $X$}\right),
\label{eq:2.3a}
 \\
 \Pi \left(E,\mbox{\boldmath $X$}\right)
= {\mathrm Im}\ {\mathrm Tr}\left\{\hat{\tau}_{l}
\frac{\partial \hat{\cal S}(E, \mbox{\boldmath $X$}
)}{\partial X_2}
\frac{\partial \hat{\cal S}^\dagger (E, \mbox{\boldmath $X$}
)}{\partial X_1}
\right\},
\label{eq:2.3b}
\end{eqnarray}
where $\mbox{\boldmath $X$}=\left(X_1,\ X_2\right)$, and $\int_A$
denotes the integration within the area encompassed by contour $A$.
It follows from the unitarity of the $S$- matrix (charge conservation)
that $\Pi \left(E,\mbox{\boldmath $X$}\right)$ 
can also be presented as
\begin{equation}
\Pi \left(E,\mbox{\boldmath $X$}\right)
= - {\mathrm Im}\ {\mathrm Tr}\left\{\hat{\tau}_{r}
\frac{\partial \hat{\cal S}(E, \mbox{\boldmath $X$}
)}{\partial X_2}
\frac{\partial \hat{\cal S}^\dagger (E, \mbox{\boldmath $X$}
)}{\partial X_1}
\right\},
\label{eq:2.3c}
\end{equation}
\end{mathletters}

Equations (\ref{eq:2.3}) are quite general. They assume only the
absence of  inelastic processes within the dot. 
We demonstrate the equivalence between 
the equations (\ref{eq:2.3}) and approach of
Ref.~\onlinecite{Zhou99} in Appendix~\ref{app1}.
 
Closing this subsection, we discuss an important point --- the
physical meaning of the adiabatic approximation necessary for
Eqs.~(\ref{eq:2.3}) to be valid.  Usually, defining a perturbation of
a quantum system as an adiabatic one means that the frequency $\omega$ of
this perturbation is much smaller than the energy of the lowest
excitations in this system. If the system were closed as in
Fig.~\ref{Fig10}, the charge distribution after each period of the
perturbation would return to the original distribution, and therefore,
the pumped charge would be {\em exactly quantized} in units of the
electron charge.

 \narrowtext {\begin{figure}[ht]
\vspace{0.2cm}
\epsfxsize=6cm
\hspace*{0.5cm}
\centerline{\epsfbox{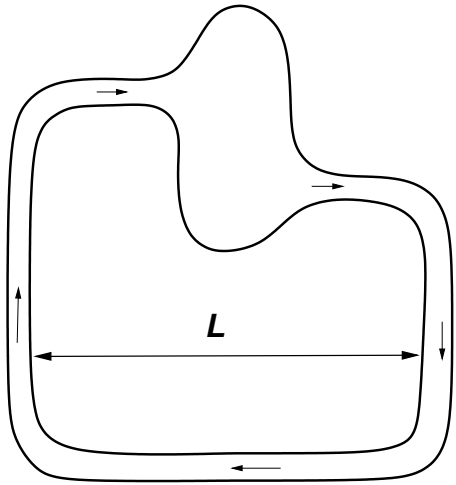}}
\vspace*{1cm}
\caption{Schematic picture of the quantum dot from Fig.~\protect\ref{Fig1} in
the closed geometry. Genuine adiabatic approximation corresponds to 
$\protect\omega \cdot L \to 0$, whereas Eqs.~(\protect\ref{eq:2.3})
assume $\protect\omega \cdot L \to \infty$.
}
\label{Fig10}
\end{figure} 
}

However, for an open system, the spectrum is continuous, and the
aforementioned adiabaticity criterium can be never  satisfied.
In this case we  the only conditionwe can impose is smallness of the frequency
as compared with the temperature and the mean level
spacing $\delta_1$ of the dot separated from the leads. 
This condition is not sufficient for the qunatization of the charge.
 Nevertheless, the value of the pumped charge can be still
expressed in the form of the adiabatic curvature (\ref{eq:2.3}).

To understand better the relation between closed and open systems,
let us consider the gedanken experiment,
where the two leadss are connected to each other as in Fig.~\ref{Fig10}.
Level spacing of the whole system $\delta_{sys}$ is proportional to
$1/L$. 
We will show now, that the result (\ref{eq:2.3}) can be decomposed into two
contributions\cite{AleinerAndreev}.
The
first contribution is {\em quantized} and is just a charge which is
pumped in truly adiabatic situation $\omega \ll \delta_{sys}$. This
charge is, therefore, attributed to the dissipationless current. The
second contribution is associated with the creation of the real
electron-hole pairs in the system and  is {\em not quantized}. It 
takes place in such closed system only provided that $\omega \gg
\delta_{sys}$. This contribution involves dissipative conductance and
it is due to the fact that the electronic system can not fully adjust
itself to the time evolution of the Hamiltonian: the state of the
system at a given time is not an eigenstate of the Hamiltonian. Such
retardation of the electrons from the external field leads to the
dissipation. (Debye losses in closed systems with $\omega \gg
\delta_1$ gives a good example of such a dissipation).

To illustrate such decomposition, we consider a simple case where each
lead contains only one channel and there is no electron-electron
interaction. In this particular case the ${\cal S}$ - matrix  is
$2\times 2$ matrix which can be parameterized in terms of the
dimensionless conductance $g$ and the scattering phase $\theta$:
\begin{equation}
\hat{\cal S}= \pmatrix{
\sqrt{1-g}e^{i\theta} & i \sqrt{g}\cr
i \sqrt{g} & \sqrt{1-g}e^{-i\theta}
}.
\label{eq:2.300}
\end{equation} 
In the presence of the pumping perturbation, both parameters $g$ and
$\theta$ are slow functions of time $t$. With the help of
Eq.~(\ref{eq:2.300}), equation (\ref{eq:2.3})
gives
\begin{equation}
Q = e\omega\int_0^{2\pi/\omega}dt \left[1-g(t)\right]\frac{d\theta}{dt}.
\label{eq:2.301}
\end{equation}
[This result was obtained by different means in Ref.~\onlinecite{AleinerAndreev}.]
One can see that the first term in brackets of the integrand
in Eq.~(\ref{eq:2.301}) gives always the quantized contribution because
$\theta\left(1/2\pi\omega\right) =\theta\left(0\right) +2\pi n $, with $n$
being an integer. At the same time, the term proportional to $g(t)$
violates this quantization, and in the case of perfect conductance
$g=1$ cancels the adiabatic contribution completely.

\subsection{Zero-dimensional model for the quantum dots}
In what follows we will be interested in the statistics of the pumped
charge in the ensemble of quantum dots. In order to determine these
statistics, we need to specify the model.  Firstly, we assume that the
size of the quantum dots $L$ is so small, that the Thouless energy
$E_T\sim \hbar/\tau_{erg}$ far exceeds other energy scales of the
problem, such as the dephasing or escape rates (here $\tau_{erg}$ is
the characteristic time for the classical particle to cover all of the
available phase space).  In this limit one can use the Random Matrix
Theory to study the conductance of the system, see
Ref.~\onlinecite{Beenakker}. All corrections to the RMT are small as
$N_{ch}/g_{dot}$, where $g_{dot}=E_T/\delta_1$ and $\delta_1$ is the
mean level spacing.  Secondly, we consider the adiabatic pumping in
the dot with the large number $N_{ch}$ of open channels. In this
approximation the effect of Coulomb interaction among the electrons in
the dot turns out to be small as $1/N_{ch}^2$ (see
Ref.\onlinecite{BrouwerAleiner}) and can be neglected.  The condition
$N_{ch}\gg 1$ also allows us to use conventional diagrammatic
technique\cite{AGD} to perform the ensemble average, and to consider only
lowest moments of the distribution of $Q$.

The Hamiltonian of the system can be represented as:
\begin{equation}
\label{eq:2.4}
\hat H=\hat H_D+\hat H_L+\hat H_{LD},
\end{equation}
where $\hat H_D$ is the Hamiltonian of electrons in the dot,
mimicked by a $M\times M$ matrix
\begin{equation}
\label{eq:2.5}
\hat H_D=\sum_{n,m=1}^M \Psi^\dag_n H_{nm} \Psi_m,
\end{equation}
We assume the ``thermodynamic''
 limit $M \sim g_{dot}\to\infty$.
In the absence of pumping perturbations $H_{nm}$ can be considered
as a random (since $g_{dot}\gg 1$)
 matrix that belongs 
to the ensemble of real symmetric matrices 
(orthogonal ensemble).
Let the perturbations be represented by two given 
not necessarily random (compare 
with Ref.~\onlinecite{SimonsAltshuler})
 $M\times M$ symmetric matrices
$V_{n,m}^{(1,2)}$, so that 
\begin{equation}
\label{eq:2.6}
H_{nm}(\mbox{\boldmath $X$}) ={\cal H}_{nm}+ 
\mbox{\boldmath $X$} \mbox{\boldmath $V$}_{nm}
={\cal H}_{nm}+X_1 V_{nm}^{(1)}
+X_2 V_{nm}^{(2)}.
\end{equation}

According to RMT, the correlation function of the
 matrix elements of the unperturbed part of the
Hamiltonian, $\hat{H}_D$,
can be written as
\begin{equation} 
\label{eq:2.7}
\langle{\cal H}_{nm}{\cal H}^*_{n'm'}\rangle=
M\left(\frac{\delta_1}{\pi}\right)^2
\left(\delta_{nn'}\delta_{mm'}
+\delta_{mn'}\delta_{nm'}\right),
\end{equation}

The coupling between the dot and the leads is
\begin{equation}
\label{eq:2.8}
\hat H_{LD}=\sum_{\alpha, n, k}\left( W_{n \alpha}\psi^\dag_\alpha
(k)
\Psi_n+
{\rm H.c.}\right),
\end{equation}
where $\Psi_n$ correspond to the states of the dot,
$\psi_{\alpha}(k)$ denotes different electron states in the leads
(momentum $k$ labels continuous spectrum
 in each channel $\alpha$).

The  spectrum of electrons
in the leads near Fermi
surface can be linearized.
 Thus, without losing the generality
we can write $\hat{H}_L$ as
\begin{equation}
\label{eq:2.9}
\hat H_L=v_F\sum_{\alpha, k} k \psi^\dag_\alpha(k)\psi_\alpha(k),
\end{equation}
where  $v_F=1/2\pi \nu$ is the
Fermi velocity and $\nu$ is the density of states at the Fermi surface.

The coupling constants $W_{n \alpha}$ in Eq.~(\ref{eq:2.8}) are 
defined in the case of the reflectionless contacts
as\cite{Beenakker}:
\begin{equation}
\label{eq:2.10}
W_{n \alpha}=\sqrt{\frac{M\delta_1}{\pi^2\nu}}
\times\cases{
1, & if $n=\alpha\leq N_{ch}$,\cr
0, & otherwise,
}
\end{equation}

For the system described above the scattering matrix 
$\hat {\cal S}$ has the form:
\begin{equation}
\label{eq:2.11}
{\cal S}_{\alpha\beta}(E,\mbox{\boldmath $X$})=1-2\pi i\nu 
W^\dag_{\alpha n} G_{nm}^R(E,\mbox{\boldmath $X$})
 W_{m\beta},
\end{equation}
and the retarded (advanced)  Green function $G_{nm}^{R}$(
 $G_{nm}^{A}$)
 is to be determined from the equation
\begin{equation}
\label{eq:2.12}
\left(E -{\hat H}(\mbox{\boldmath $X$})\pm
i\pi\nu \hat{W}\hat{W}^\dag \right)\hat{G}^{R,A}(E,\mbox{\boldmath $X$})=
\hat{I}.
\end{equation}
Here matrices $\hat{H}$ and $\hat{W}$ are
 comprised by their elements (\ref{eq:2.6}) and (\ref{eq:2.10}) respectively.
 The factor in Eq.~(\ref{eq:2.10}) is chosen so that
the ensemble average scattering matrix $\langle{\cal
S}_{\alpha\beta}\rangle$ of a
dot with fully open channels
is zero. More complicated structure of $\hat W$ can be
always reduced to the form (\ref{eq:2.10}) by suitable rotations. 
We are going to consider the averages of different moments of
$\Pi$ given by Eq.~(\ref{eq:2.3b}) with the help of diagrammatic
technique\cite{AGD}. For the technical reasons it is more
convenient to transform Eq.~(\ref{eq:2.3b}) with the help
of Eq.~(\ref{eq:2.3c}):
\begin{eqnarray}
&&\Pi \left(E,\mbox{\boldmath $X$}\right)
= {\mathrm Im}\ {\mathrm Tr}\left\{\hat{\tau}
\frac{\partial \hat{\cal S}(E, \mbox{\boldmath $X$}
)}{\partial X_2}
\frac{\partial \hat{\cal S}^\dagger (E, \mbox{\boldmath $X$}
)}{\partial X_1}
\right\},\nonumber \\
&&\hat{\tau}=
\frac{N_r \hat{\tau}_l - N_l \hat{\tau}_r}{N_{ch}} 
\label{eq:2.13}
\end{eqnarray}
Notice that the  matrix $\hat{\tau}$ is traceless.
 This fact significantly simplifies
further manipulations. 
We substitute  Eq.~(\ref{eq:2.11}) into
Eq.~(\ref{eq:2.13}) and obtain
\begin{eqnarray}
&&\Pi \left(E,\mbox{\boldmath $X$}\right)
=-\frac{i}{2} 
\varepsilon^{ij}
\frac{\partial^2{\cal F}}{\partial X_i^A \partial X_j^R}
{\Big |}_{\mbox{\boldmath
$X$}^R=\mbox{\boldmath $X$}^A=\mbox{\boldmath $X$}}
,\nonumber\\
&&{\cal F}=s\frac{4M^2\delta_1^2}{\pi^2}
{\mathrm Tr}\left\{\hat{\Lambda}\hat{G}^R(E,\mbox{\boldmath
$X$}^R)\hat{\Gamma}
\hat{G}^A(E,\mbox{\boldmath
$X$}^A)
\right\},\label{eq:2.14}
\end{eqnarray}
where $\varepsilon^{ij} =\left[\hat{\varepsilon}\right]^{ij} 
$ is the antisymmetric tensor of the second rank, and 
the factor $s=2$ takes into
account the spin degeneracy.
We have introduced matrices $\hat{\Lambda}$
and $\hat{\Gamma}$:
\begin{eqnarray}
\left[\hat{\Gamma}\right]_{nm}
&=&\delta_{nm}\times
\left\{\matrix{1, & 1 \leq n \leq N_{ch}\cr
0, & N_{ch} < n \leq M
}\right.\label{eq:2.15}
\label{matrices}
\\ 
\left[\hat{\Lambda}\right]_{nm}
&=&\delta_{nm}\times
\left\{\matrix{\frac{N_r}{N_{ch}}, & 1 \leq n \leq N_{l}\cr
-\frac{N_l}{N_{ch}}, & N_{l} < n \leq N_{ch}\cr
0, & N_{ch} < n \leq M
}\right.
 \nonumber
\end{eqnarray}
Notice, that 
\begin{equation}
\label{eq:2.150}
{\mathrm Tr}\hat\Lambda= 0, \quad {\mathrm Tr}\left(
\hat\Lambda\hat\Gamma\right)=0.
\end{equation}

To evaluate correlators  of the functions
Eq.~(\ref{eq:2.14}), we adopted a diagrammatic technique
for the ensemble  averaging.   
 In the thermodynamic limit $M \to \infty$ 
this technique is somewhat similar to the one
 developed for 
bulk disordered metals \cite{AGD}. Factor $1/M$
plays now the same role as  the small parameter
$1/\epsilon_F\tau_{imp}$ with $\epsilon_F$ and $\tau_{imp}$ 
being the Fermi energy 
and the elastic mean free time 
correspondingly. The rules for reading those diagrams
are shown in Fig.~\ref{Fig2}a.

\narrowtext
{\begin{figure}[ht]
\vspace{0.2cm}
\epsfxsize=6cm
\hspace*{0.5cm}
\epsfbox{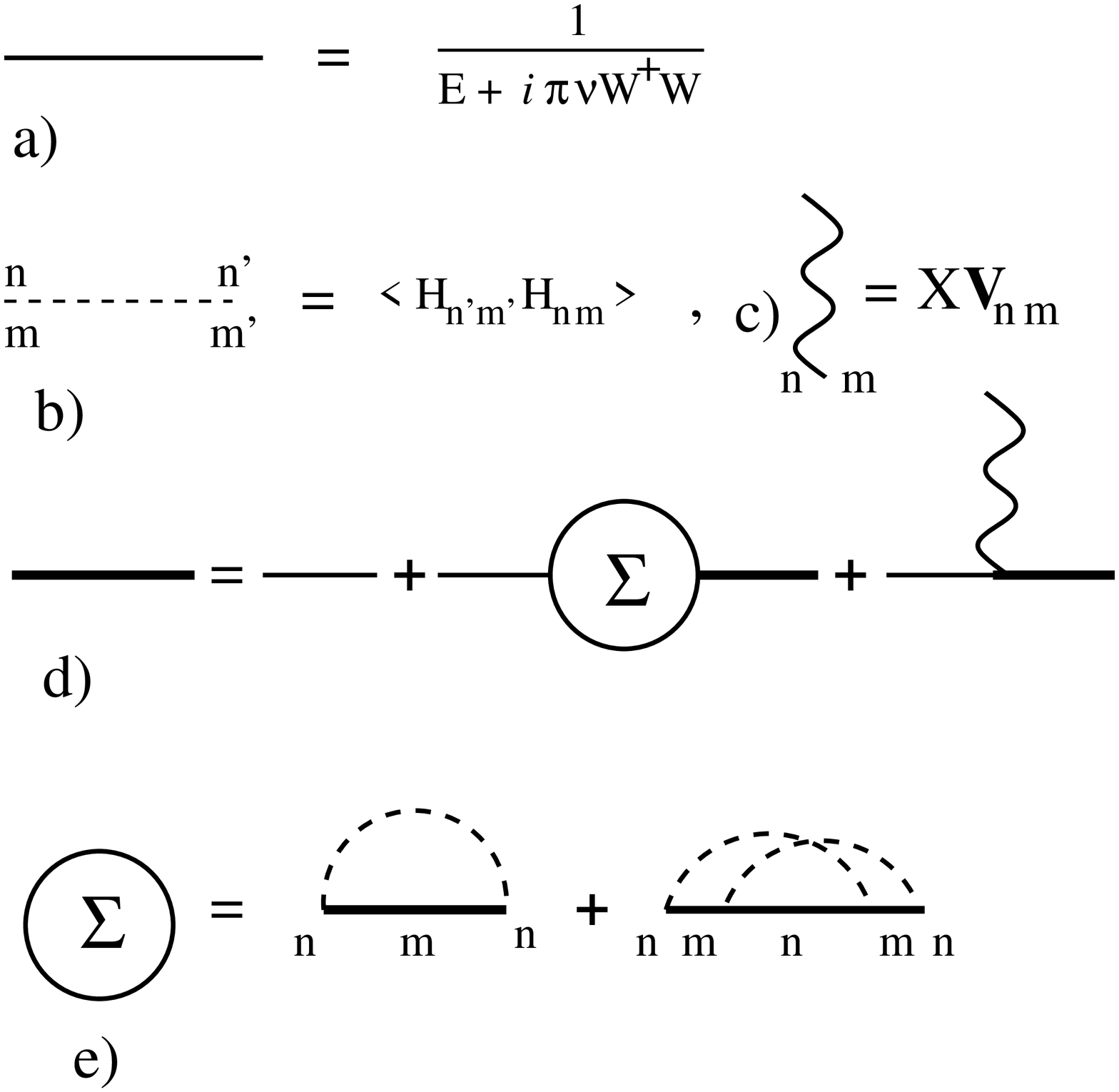}
\vspace*{0.5cm}
\vspace{0.9 cm}
\caption{Elements of the diagram technique:
(a) bare electron retarded Green function;
(b) correlator of the matrix elements of the Hamiltonian;
(c) pumping perturbation;
(d) renormalized electron Green function; 
(e) self-energy $\hat{\Sigma}$. The second term in the
self-energy,
which includes an intersection of the dashed lines,  is
smaller compared to the first term as $1/M$.
}
\label{Fig2}
\end{figure} 
}
The ensemble averaged retarded Green function 
is given by
\begin{eqnarray}
&&\hat{\cal G}^{R,A}  ={\Big \langle}
\frac{1}{E -\hat{H}(\mbox{\boldmath $X$})\pm
i\hat{\Gamma}{\cal E}
}{\Big \rangle} = \nonumber\\
&&=\frac{1}
{E -\hat{H}(\mbox{\boldmath $X$})\pm
i\hat{\Gamma}{\cal E} 
- \hat{\Sigma}^{R,A}}.
\label{eq:2.16}\\
&& {\cal E}=\frac{M\delta_1}{\pi}\nonumber
\end{eqnarray} 
where averaging is performed over realizations of random matrix
$\hat{\cal H}$ from Eq.~(\ref{eq:2.6}), and energy ${\cal
E}$ is of the same order as the width of the band of the
random matrix eigenvalues.
Self energy $\hat{\Sigma}$ includes, as usual,  all of the one-particle irreducible
graphs. In the leading in $1/M$ approximation it is given by the sum
of the rainbow diagrams,  Fig.~\ref{Fig2}b,
\begin{equation}
\hat{\Sigma}^{R,A} = M\left(\frac{\delta_1}{\pi}\right)^2
\hat{I}  {\mathrm Tr} \left\{
\hat{\cal G}^{R,A}
\right\}.
\label{eq:2.160}
\end{equation}
Let us now
 expand the {\em ensemble averaged} Green function up to the second
order in parametric perturbation. 
The final results are not necessarily quadratic 
in perturbation strength, see e.g., Eq.~(\ref{eq:2.19}).
All higher order terms
can be neglected provided that
\begin{equation}
|V_{nm}^{(i)}| \ll M\left(\frac{\delta_1}{\pi}\right) 
, \quad n,m = 1,2,\dots,M.
\label{eq:2.22}
\end{equation}
Inequality (\ref{eq:2.22}) is
nothing but the condition 
of  the central limit theorem.
Solving Eqs.~(\ref{eq:2.16}) and (\ref{eq:2.160}), we find 
that at $E \ll
{\cal E}$ 
\begin{eqnarray}
\label{eq:2.17}
&&\hat{\cal G}^{R,A}=
\pm\frac{1}{i{\cal E}}
\left[\hat{I} - \frac{1}{2} \hat{\Gamma}
+ \frac{N_{ch}\pm i\epsilon}{4M}\hat{I}
\right]-\\
&&-
\frac{1}{{\cal E}^2}
\left(
\mbox{\boldmath $X$}\mbox{\boldmath $V$}\right)
\left(\hat{I}\pm\frac{1}{i{\cal E}}
\mbox{\boldmath $X$}\mbox{\boldmath $V$}\right)
+\nonumber\\
&&+\frac{1}{{\cal E}^2}\frac{\hat{I}}{2M}
{\mathrm Tr}
\left(\mbox{\boldmath $X$}\mbox{\boldmath $V$}
\pm
\frac{1}{i{\cal E}}\left(\mbox{\boldmath $X$}\mbox{\boldmath $V$}
\right)
\left(\mbox{\boldmath $X$}\mbox{\boldmath $V$}\right)\right),
\nonumber
\end{eqnarray}
where $\hat{\Gamma}$ is defined in Eq.~(\ref{matrices}).
Here we introduced the dimensionless energy measured in units of mean
level spacing 
\begin{equation}
\epsilon = \frac{2\pi E}{\delta_1}.
\label{eq:2.170}
\end{equation}
 We can also expand these Green 
functions up to 
the first order in $\epsilon/M$ and $N_{ch}/M$
and restrict ourselves by zero and first order terms,
 since the higher order terms 
vanish in the ``thermodynamic'' limit
$M\to\infty$. 

Let us now turn to the analysis of the statistical properties of
the function ${\cal F}$ from Eq.~(\ref{eq:2.14}). 
All the relevant
averages will be expressed in terms of the certain products of the
Green functions -- diffuson ${\cal D}$ and Cooperon ${\cal C}$
\cite{AltshulerAronov}:
\begin{mathletters}
\label{eq:2.18}
\begin{eqnarray}
{\cal D}
\left(\epsilon, \mbox{\boldmath $X$}\right) = 
{\mathrm Tr}
\langle 
\hat{G}^{R}\left(\epsilon_1+\epsilon, 
\mbox{\boldmath $X$} + \mbox{\boldmath $Y$}\right)
\hat{G}^{A}\left(\epsilon_1, \mbox{\boldmath $Y$}\right) 
\rangle
 \label{eq:2.18a}\\
{\cal C}
\left(\epsilon, \mbox{\boldmath $X$}\right) =
{\mathrm Tr}
\langle 
\hat{G}^{R}\left(\epsilon_1+\epsilon, 
\mbox{\boldmath $X$} + \mbox{\boldmath $Y$}\right)
\hat{G}^{A}\left(\epsilon_1, \mbox{\boldmath $Y$}\right)
^T 
\rangle,
 \label{eq:2.18b}
\end{eqnarray}
\end{mathletters}
where we express energy in dimensionless units (\ref{eq:2.170}).

The leading  at $M \to \infty$ and $N_{ch} \gg 1$
approximation for the diffuson  is a series of ladder diagrams.
As usual,
summation of this 
series can be performed by solving the equations presented 
graphically on 
 Fig.~\ref{Fig3}. The solution of this diagrammatic equation
is
\begin{equation}
\label{eq:2.19}
{\cal D}={\cal C}
=\left(\frac{2\pi}{\delta_1}\right)^2
\frac{1}{-i\epsilon + N_{ch} + 
i\mbox{\boldmath $Z$}\cdot \mbox{\boldmath $X$}
+\mbox{\boldmath $X$}\hat{C}_0\mbox{\boldmath $X$}
}.
\end{equation}
The diffuson has the universal form  
Eq.~(\ref{eq:2.19}) under the  condition  of 
Eq.~(\ref{eq:2.22}).
The structure and the strength of the perturbation potential 
$\hat{\mbox{\boldmath $V$}}$ from Eq.~(\ref{eq:2.6}) is encoded in two
parameters: vector $\mbox{\boldmath $Z$}$ and tensor $\hat{C}_0$. In
terms of the original Hamiltonian they are given by
\begin{mathletters}
\label{eq:2.20}
\begin{eqnarray}
\label{eq:2.20a}
\mbox{\boldmath $Z$} &=& \frac{2\pi}{M\delta_1}{\mathrm Tr}
\hat{\mbox{\boldmath $V$}}\\
\label{eq:2.20b}
\left[\hat{C}_0\right]_{ij}&=&\frac{2\pi^2}
{(M\delta_1)^2} 
{\mathrm Tr}\left\{\hat{V}^{(i)}\hat{V}^{(j)}\right\}
, \ i,j =1,2,
\end{eqnarray}
We used the fact that the matrix $\hat{V}$ is symmetric.  
Parameters (\ref{eq:2.20})
 are  also related to the typical value of the level
velocities, which characterizes the evolution of energy levels of the
closed system $\epsilon_{\nu}(\mbox{\boldmath $X$})$ under 
the action of an external perturbation 
$\mbox{\boldmath $X$}\cdot \hat{\mbox{\boldmath $V$}}$, 
see Ref.~\onlinecite{SimonsAltshuler} 
(our definition is different by a numerical factor):
\end{mathletters}
\begin{mathletters}
\label{eq:2.21}
\begin{eqnarray}
\mbox{\boldmath $Z$}&=& \left(\frac{2\pi}{\delta_1}\right)
 {\Big\langle}
\frac{\partial
\epsilon_{\nu}}{\partial \mbox{\boldmath $X$}} {\Big\rangle},
\label{eq:2.21b}
\\
\left[\hat{C}_0\right]_{ij}&=& 
\frac{\pi^2}{\delta_1^2}
\left(
{\Big\langle} 
\frac{\partial \epsilon_\nu}{\partial
X_i} \frac{\partial \epsilon_{\nu}}{\partial X_j}{\Big\rangle} - 
{\Big\langle} \frac{\partial
\epsilon_{\nu}}{\partial X_i} {\Big\rangle}{\Big\langle} \frac{\partial
\epsilon_{\nu}}{\partial X_j} {\Big\rangle}
\right).  
\end{eqnarray} 
Now we are in a state to evaluate the statistics
of the pumped charge. It should be noted that
the specifics of the system enter {\em only}
through the parameters $Z_i$ and $[\hat{C}_0]_{ij}$.
 Moreover,
 all other responses of the system 
(e.g. parametric dependence of the
conductance of the dot) are also universal functions
of the same parameters\cite{SimonsAltshuler}. Therefore, 
$Z_i$ and $[\hat{C}_0]_{ij}$
 can be in principle determined
from
independent measurements. 
\end{mathletters}

\narrowtext
{\begin{figure}[ht]
\vspace{0.2cm}
\epsfxsize=6cm
\hspace*{0.5cm}
\epsfbox{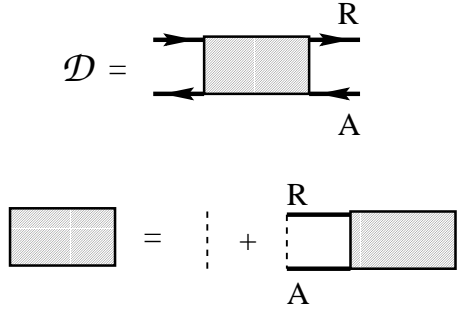}
\vspace*{0.5cm}
\vspace{0.9 cm}
\caption{Diffuson  diagrams.
Cooperon diagrams differ from the ones 
for diffuson only in the direction of
the arrow of the retarded Green function line.
}
\label{Fig3}
\end{figure} 
}

The ensemble average of the pumped charge vanishes and
 the sign of 
 $Q$  is random. To evaluate the typical value
of this charge we consider the average $\langle Q_AQ_B\rangle$ for
two different contours $A$ and $B$ 
on the parameter planes. To accomplish 
this task, we need to average the 
product of functions (\ref{eq:2.14})
\begin{equation}
{\cal I} = 
{\Big \langle} 
{\cal F}(\epsilon_1 +\epsilon,\mbox{\boldmath
$X$}^R, \mbox{\boldmath
$X$}^A){\cal F}(\epsilon_1,\mbox{\boldmath
$Y$}^R, \mbox{\boldmath
$Y$}^A)
{\Big \rangle},
\label{eq:2.23}
\end{equation}
where function ${\cal F}$ is defined by Eq.~(\ref{eq:2.14}).

Diagrammatic expression for ${\cal I}$ is shown on
Fig.~\ref{Fig4}. Due to the relations (\ref{eq:2.150}), 
there is no need to renormalize the vertex
$\Lambda$ by the dashed lines. For the same
reason, vertices $\Lambda$ and $\Gamma$ can not appear
in the same cell
and, therefore, the Cooperon Eq.~(\ref{eq:2.18b})
does not contribute 
to ${\cal I}$
(\ref{eq:2.23}). We demonstrate
in Sec.~\ref{sec:3}, that this
implies the difference in how the change of sign of 
magnetic field affects the pumped charge and the conductance
\cite{Onsager,Bonsager,AltshulerSpivak}.
\narrowtext
{\begin{figure}[ht]
\vspace{0.2cm}
\epsfxsize=6cm
\hspace*{0.5cm}
\epsfbox{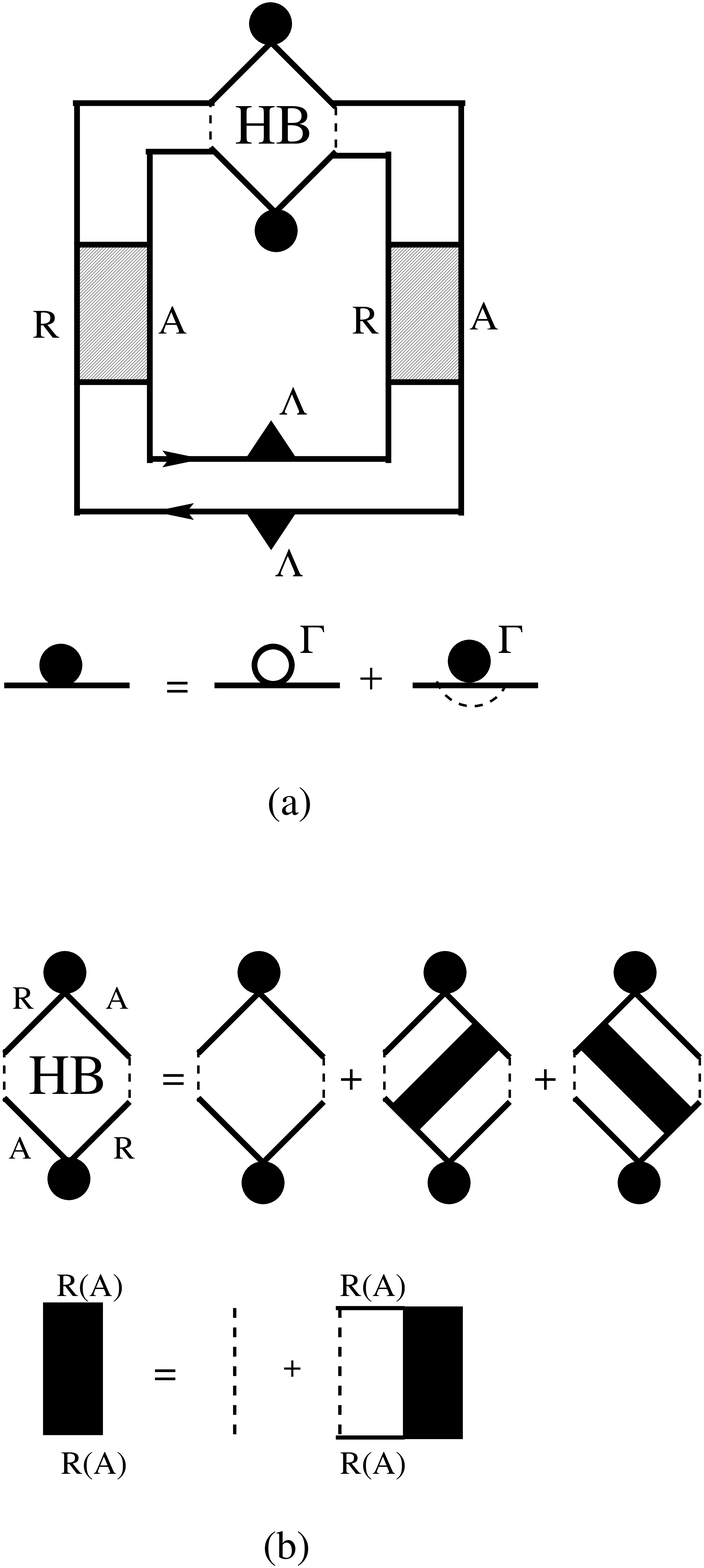}
\vspace*{1cm}
\vspace{0.9 cm}
\caption{(a) Diagrammatic representation for 
the correlation function ${\cal I}$ from
Eq.~(\protect\ref{eq:2.23}); (b) Hikami box.
}
\label{Fig4}
\end{figure} 
}

The analytic expression for the diagram, Fig.~\ref{Fig4}, is
\begin{equation}
{\cal I} = \frac{N_lN_r}{N_{ch}}{\cal B}
{\cal D}\left(\epsilon;\mbox{\boldmath $X$}^R 
- \mbox{\boldmath $Y$}^A\right)
{\cal D}\left(-\epsilon;\mbox{\boldmath $Y$}^R 
- \mbox{\boldmath $X$}^A\right),
\label{eq:2.24}
\end{equation}
where diffuson propagator ${\cal D}$ is given by Eq.~(\ref{eq:2.19}).
The factor ${\cal B}$ in Eq.~(\ref{eq:2.24}) is given by
the set of diagrams Fig.~\ref{Fig4}b, which is
 analogous to the Hikami
box for the disordered systems. It equals to
\begin{eqnarray}
{\cal B}&=& {\cal B}^{(1)} +
 {\cal B}^{(2)} + {\cal B}^{(3)}; 
\label{eq:2.25}\\
 {\cal B}^{(1)} &=&
\frac{\delta_1^4}{8\pi^4}
\left[
\left(\mbox{\boldmath $X$}^R - \mbox{\boldmath $X$}^A\right)
\hat{C}_0
\left(\mbox{\boldmath $Y$}^A - \mbox{\boldmath $Y$}^R\right)
\right];
 \nonumber\\
{\cal B}^{(2)} &=& \frac{\delta_1^2}
{4 \pi^2{\cal D}\left(\epsilon;\mbox{\boldmath $X$}^R 
- \mbox{\boldmath $Y$}^A\right)}
; \nonumber\\
{\cal B}^{(3)} &=& \frac{\delta_1^2}
{4 \pi^2{\cal D}\left( - \epsilon;\mbox{\boldmath $Y$}^R 
- \mbox{\boldmath $X$}^A\right)}.
\nonumber
\end{eqnarray}
Substitution of the expression for ${\cal B}^{(2)}$ into 
Eq.~(\ref{eq:2.24}) 
gives no contribution to ${\cal I}$
due to the relationship
between ${\cal I}$ and $\Pi$, 
Eq.~(\ref{eq:2.14}). Indeed,  the part of  ${\cal I}$ 
which depends on  ${\cal B}^{(2)}$ does not
contain  $\mbox{\boldmath $X$}^R$.
 The contribution
proportional to ${\cal B}^{(3)}$
vanishes in a similar way. 
Substituting ${\cal B}^{(1)}$ into Eq.~(\ref{eq:2.14}),
and the result into Eq.~(\ref{eq:2.24}), we find
\begin{eqnarray}
{\Big \langle}
&\Pi& \left(\epsilon_1, 
\mbox{\boldmath $X$}+ \mbox{\boldmath
$Y$}\right)
\Pi \left(\epsilon_2, 
\mbox{\boldmath $Y$}\right)
{\Big \rangle}=\frac{8 N_lN_r}{N_{ch}}
\left[\hat{\varepsilon}\hat{C}_0\hat{\varepsilon}
\right]_{ij}\frac{\partial^2}{\partial X_i  \partial X_j}
\nonumber\\
&\times&
\frac{1}{
\left(\epsilon_1-\epsilon_2 - \mbox{\boldmath $Z$}\cdot\mbox{\boldmath
$X$}\right)^2 + \left(N_{ch} + \mbox{\boldmath $X$}
\hat{C}_0\mbox{\boldmath $X$}\right)^2,
}
\label{eq:2.26}
\end{eqnarray}
Here we used the explicit form of the diffuson (\ref{eq:2.19}),
introduced dimensionless energies (\ref{eq:2.20}) and took the spin
degeneracy into account.

Now, we are in a state to  evaluate 
the correlation function of charges pumped
in a course of  motion along  contours $A$ and $B$ on the parameter
plane. The result becomes compact if we
choose new variables:
\begin{mathletters}
\label{eq:2.27}
\begin{eqnarray}
\label{eq:2.27a}
\mbox{\boldmath $x$}&=&
\frac{1}{\sqrt{N_{ch}}} \left(\hat{C}_0\right)^{1/2}     
\mbox{\boldmath $X$}; \\
\label{eq:2.27b}
\mbox{\boldmath $y$}&=&
\frac{1}{\sqrt{N_{ch}}} \left(\hat{C}_0\right)^{1/2}     
\mbox{\boldmath $Y$}; \\
\label{eq:2.27c}
\mbox{\boldmath $z$}&=&
\frac{1}{\sqrt{N_{ch}}} \left(\hat{C}_0\right)^{- 1/2}     
\mbox{\boldmath $Z$}.
\end{eqnarray}
\end{mathletters}

Let us discuss why $\mbox{\boldmath $x$}, 
\mbox{\boldmath $y$},
\mbox{\boldmath $z$}$ are natural dimensionless variables.
Recall that we are dealing with open systems.
Electronic escape time can be estimated as 
$\tau_{esc}\sim (N_{ch}\delta_1)^{-1}$.
All the energy levels have a finite width 
$\gamma \sim 1/\tau_{esc}\sim N_{ch}\delta_1>\delta_1$.
It means that even at $T=0$ the pumping current is 
 determined by the energy strip with a finite width
 $\gamma>\delta_1$. The number of the levels $n_{\gamma}$
in such a strip is a random function of the point
in the parameter space. The correlation length 
$R=\left|\mbox{\boldmath $R$}\right|$ of this random
function can be estimated from the equation
\begin{equation}
\label{eq:2.271}
\mbox{\boldmath $R$}\mbox{\boldmath $Z$}+
R_i\left[\hat{C}_0\right]_{ij}R_j\sim\gamma
\end{equation}   
The first term in the left hand side of Eq.~(\ref{eq:2.271})
describes a homogeneous shift of the spectrum, 
while the second term represents random parametric oscillations
\cite{SimonsAltshuler}. Eq.~(\ref{eq:2.271}) can
be rewritten in terms of the new variables   
Eq.~(\ref{eq:2.27}) as
\begin{equation}
\label{eq:2.272}
\mbox{\boldmath $r$}\mbox{\boldmath $z$}+
\mbox{\boldmath $r$}^2\sim 1
\end{equation} 
where $\mbox{\boldmath $r$}=(\hat{C}_0/N_{ch})^{1/2}
\mbox{\boldmath $R$}$. As a result, 
$r\sim  \mathrm{min}(1,\tilde{Z}^{-1})$.
It means that in terms of the new variables
 $\mbox{\boldmath $x$}, 
\mbox{\boldmath $y$},
\mbox{\boldmath $z$}$ the pumping 
is weak (i.e. bilinear) provided that
$r < \mathrm{min}(1,z^{-1})$.
Finally, 
\begin{eqnarray}
\label{eq:2.28}
&&\langle Q_AQ_B\rangle
=- \frac{e^2 N_lN_r}{N_{ch}^2}\int_{a}dx_1
dx_2
\int_{b}dy_1dy_2 \\
&&\times 
\left(\frac{\partial}{\partial \mbox{\boldmath $x$}_-}\right)^2
{\cal K}
\left(
\frac{2\pi T}{N_{ch}\delta_1},
\mbox{\boldmath $z$}\cdot\mbox{\boldmath $x$}_-,
\mbox{\boldmath $x$}_-^2  
\right),\nonumber \\
&&\mbox{\boldmath $x$}_-=
\mbox{\boldmath $x$}-\mbox{\boldmath $y$},
\nonumber
\end{eqnarray}
where $\int_{a,b}$
denotes the integration within the area encompassed by contours
$a$ and $b$,  see
Eq.~(\ref{eq:2.27}), and dimensionless correlation function 
${\cal K}$ is given by
\begin{eqnarray}
\label{eq:2.29}
&&{\cal K}(u,v,w)= \frac{8}{\pi^2}\int_{-\infty}^\infty dh
 \frac{f(h)}
{ 
\left[\left(2 u h+v\right)^2 + \left(1+w\right)^2\right]}.\\
&& f(h)= \frac{(h \mathop{\mathrm{cotanh}} h -1)}
{\left(\sinh h\right)^2
}\nonumber
\end{eqnarray}
\begin{mathletters}
Low temperature regime corresponds 
to $u \ll {\mathrm max}(1,v,w)$. In this limit 
\begin{equation}
\label{eq:2.30a}
{\cal K}(u,v,w)= \frac{8}{\pi^2}
\frac{1}{v^2 + \left(1+w\right)^2}.
\end{equation}
while at high temperatures $u \gg {\mathrm max}(1,v,w)$ 
\begin{equation}
\label{eq:2.30b}
{\cal K}(u,v,w)= \frac{4}{3\pi u}
\left(\frac{1}{1+w}\right).
\end{equation}
Therefore heating suppression of the mesoscopic
fluctuations of $Q$ is similar to that
 of the conductance fluctuations. 
\end{mathletters}

Equations (\ref{eq:2.28}) -- (\ref{eq:2.29}) are the main results of
this section. They describe the correlation between the charge pumped 
due to the motion in the parameter space along the different contours
 at arbitrary temperature. Now, we are going to apply 
Eq.~(\ref{eq:2.28}) to analyze the variance of the charge 
$\langle Q^2_A\rangle$. 

\subsection{Weak Pumping.}
\label{sec:2.5}

If  the characteristic magnitude of the potentials
is so small that 
$\mbox{\boldmath $x$}_-^2 \ll 1, \  
\mbox{\boldmath $x$}_-\cdot
\mbox{\boldmath $z$} 
\ll {\mathrm max} \left(1,
T/(N_{ch}\delta_1)\right)$,  then 
the system is in bilinear response regime
discussed in Refs.~\onlinecite{Zhou99,Brouwer99}. In this case one can
put $v=0,w=0$ in Eq.~(\ref{eq:2.29}) after the differentiation.
As a result
\begin{equation}
\label{eq:2.31}
\langle Q_A^2\rangle = 
 \frac{e^2 N_lN_rS_{a}^2}{N_{ch}^2}
\left[
{\cal K}_1\left(\frac{2\pi T}{N_{ch}\delta_1}\right) +
\mbox{\boldmath $z$}^2
{\cal K}_2\left(\frac{2\pi T}{N_{ch}\delta_1}\right)
\right], 
\end{equation}
where $S_{a}$ is the area
enclosed by the contour $a$, in the
parameter space $\mbox{\boldmath $x$} = \left(x_1,x_2\right)$. Functions ${\cal
K}_{1,2}(x)$ can be  expressed through
$f(h)$ from Eq.~(\ref{eq:2.29}) in the following way: 
\begin{mathletters}
\label{eq:2.32}
\begin{eqnarray}
{\cal K}_1(u)&=& \frac{32}{\pi^2}\int_{-\infty}^\infty dh
 \frac{f(h)}
{\left[4 u^2 h^2 + 1\right]^2}
\nonumber\\
&=&
\left\{
\matrix{
\frac{32}{\pi^2}, & u \ll 1; \cr
\frac{8}{3\pi u}, & u \gg 1},
\right.
\label{eq:2.32a}
\end{eqnarray} 
and
\begin{eqnarray}
{\cal K}_2(u)&=& \frac{16}{\pi^2}
\int_{-\infty}^\infty dh
 \frac{f(h)
\left(1 - 12 u^2 h^2\right)
 }
{\left[4 u^2 h^2 + 1\right]^3}
\nonumber\\
&=&
\left\{
\matrix{
\frac{16}{\pi^2}, & u \ll 1; \cr
\frac{4}{15 \pi u^3}, & u \gg 1.}
\right.
\label{eq:2.32b}
\end{eqnarray}
\end{mathletters}
In terms of the original pumping strength 
$\mbox{\boldmath $X$}$  Eq.~(\ref{eq:2.31})
acquires the form
\begin{eqnarray}
\label{eq:2.321}
\langle Q_A^2\rangle &=& 
 \frac{e^2 N_lN_rS_{A}^2}{N_{ch}^4}
\left(\mathrm{det}\left[\hat{C}_0\right]\right)^2\times
\\
&&\left[
{\cal K}_1\left(\frac{2\pi T}{N_{ch}\delta_1}\right) +
\frac{1}{N_{ch}}Z_i
\left[\hat{C}_0^{-1}\right]^{ij}Z_j
{\cal K}_2\left(\frac{2\pi T}{N_{ch}\delta_1}\right)
\right]
\nonumber
\end{eqnarray}
Note that the specifics of the system
enter {\em only} through the vector $\mbox{\boldmath $Z$}$
and the tensor $\hat{C_0}$ defined in Eq.~(\ref{eq:2.20}).
In high temperature regime ${\cal K}_2 \ll {\cal K}_1$.
It means that the simultaneous shift of all 
levels ( determined by $\mbox{\boldmath $z$}$)
is not relevant for pumping. On the other hand, 
at $T \to 0$, this simultaneous shift of all levels
may be important.

\subsection{Strong Pumping.}
\label{sec:2.6}

Let us now turn to the discussion of the opposite limit, 
 $\mbox{\boldmath $x$}_-^2 \gg 1$, where pumping is strong. In this
regime it is more convenient to transform Eq.~(\ref{eq:2.28}) to the
contour integrals. Using Stokes theorem, we find
\begin{eqnarray}
\langle Q_A^2\rangle
&=&\frac{e^2 N_lN_r}{N_{ch}^2}
\oint_{a}dx_i\oint_{a}dy_i
{\cal K}
\left(
\frac{2\pi T}{N_{ch}\delta_1},
\mbox{\boldmath $z$}\mbox{\boldmath $x$}_-,
\mbox{\boldmath $x$}_-^2  
\right),\nonumber \\
\mbox{\boldmath $x$}_-&=&
\mbox{\boldmath $x$}-\mbox{\boldmath $y$},
\label{eq:2.33}
\end{eqnarray}
We notice from Eq.~(\ref{eq:2.29}) that the kernel ${\cal K}$ decreases
rapidly at $\mbox{\boldmath $x$}_-^2 \gtrsim 1$. Since the
characteristic scale of the field itself is large
$\mbox{\boldmath $x$}^2 \gg 1$, we can perform the integration
over $\mbox{\boldmath $x$}_-$ locally along the direction of
the contour $d\mbox{\boldmath $x$}$. It gives
\begin{equation}
\langle Q_A^2\rangle
= \frac{e^2 N_lN_r}{N_{ch}^2}
\oint_{a}dx
{\cal L}
\left(
\frac{2\pi T}{N_{ch}\delta_1},
\frac{\mbox{\boldmath $z$} \times \mbox{\boldmath
$x$}}
{\left|
\mbox{\boldmath
$x$}
\right|}
\right).
\label{eq:2.34}
\end{equation}
The kernel ${\cal L}$ can be expressed in terms of ${\cal K}$ from
Eq.~(\ref{eq:2.29}) as
\begin{equation}
{\cal L}
\left(u,v\right)
=\int_{-\infty}^{\infty}dw {\cal K}(u,vw,w^2).
\label{eq:2.35}
\end{equation}

If the averaged level velocity is small, $y \ll 1$, we find from 
Eqs.~(\ref{eq:2.29}) and Eqs.~(\ref{eq:2.34})
\begin{equation}
\langle Q_A^2\rangle
= e^2 {\ell}_a \frac{ N_lN_r}{N_{ch}^2}
{\cal L}_1
\left(
\frac{2\pi T}{N_{ch}\delta_1}
\right),
\label{eq:2.36}
\end{equation}
where ${\ell}_a$ is the {\em length} of the contour
$a$.
Function ${\cal L}_1$ in Eq.~(\ref{eq:2.36}) is given by
\begin{eqnarray}
{\cal L}_1(u)&=& \frac{2^{5/2}}{\pi}\int
 \frac{ dh f(h) }
{ \left[1+ 4 u^2 h^2 + \left(1+ 4 u^2 h^2\right)^{3/2}\right]^{1/2}}
\nonumber\\
&=&
\matrix{
\frac{4}{\pi}, & u \ll 1 \cr
\frac{4}{3 u}, & u \gg 1}
\label{eq:2.37}
\end{eqnarray}
In terms of the original pumping strength
$\mbox{\boldmath $X$}$,
the dimensionless length of the contour
${\ell}_a$ ( Eq.~(\ref{eq:2.36}))
acquires the form
\begin{equation}
{\ell}_a
=  \frac{1}{N_{ch}}
\oint_A\sqrt{dX_i\hat{C}^{ij}_{0}dX_j}
\label{eq:2.361}
\end{equation}

 It is important to emphasize that in the case of the
strong perturbation, the pumped charge is determined by the length of
the contour rather than by its area, and it is not sensitive to the
contour shape  
(provided that the contour 
is smooth on the scale of the order of unity). 
It  has to be contrasted with the naive expectation  
$\langle Q_A^2\rangle \propto S_a$, which follows
from independent addition of areas.

If  
$\mbox{\boldmath $z$}$
is not small, the value of the pumped
charge depends not only on the length of the contour but also on its
shape. At low temperatures, we obtain from Eqs.~(\ref{eq:2.34}),
(\ref{eq:2.35}), and (\ref{eq:2.30a})
\begin{equation}
\langle Q_A^2\rangle
= \frac{8 e^2 N_lN_r}{\pi N_{ch}^2}
\oint_{a}dx
\frac{1}{
\left[
4 + \mbox{\boldmath $z$}^2 \sin^2 
\left(
\widehat{\mbox{\boldmath $z$}\mbox{\boldmath
$x$}} 
\right)
\right]^{1/2}}
,
\label{eq:2.38}
\end{equation}
where $\widehat{\mbox{\boldmath $z$}\mbox{\boldmath
$x$}}$ is an angle between the vectors $\mbox{\boldmath $z$}$
and $\mbox{\boldmath $x$}$.
At high temperature, $\langle Q^2\rangle $ does not depend on
on the shape of the contour and  is determined by  
Eqs.~(\ref{eq:2.36}) and (\ref{eq:2.37}).

\section{Magnetic field effects on adiabatic pumping.}
\label{sec:3}
This section is devoted to the effect of the magnetic field on the
pumped charge. In subsection ~\ref{sec:3.2}, we present a general
discussion of the symmetries with respect to the time inversion.
We demonstrate
that, unlike the conductance, the pumped charge does not possess such
a symmetry. This general conclusion is illustrated in subsection
~\ref{sec:3.2} by the model calculation of the
second moment of the charge
pumped through the quantum dot.

\subsection{Symmetry with respect to the reversal of the magnetic field}
\label{sec:3.1}

Let us now consider the pumping through the mesoscopic sample
subjected to a magnetic field $B$. The general formalism of
Sec.~\ref{sec:2.1} remains valid. One can  infer from
Eqs.~(\ref{eq:2.3}) that the sign of the pumped charge 
changes together with the direction of the contour
in the parameter space
\begin{equation}
Q_{\hookrightarrow}(B) =- Q_{\hookleftarrow}(B)
\label{eq:3.1}
\end{equation}
where ${\hookrightarrow}({\hookleftarrow})$ denote opposite direction
of motion in the parameter space along the same contour.  
Indeed, curvature (\ref{eq:2.3b}), is a single valued function of its
parameters $X_1,\ X_2$. Therefore, the only effect of reversal of
the contour direction is to change the sign of the directed area
$dX_1dX_2 \to -dX_1dX_2$ without changing the integration domain. This
immediately yields identity (\ref{eq:3.1}).
Note that Eq.~(\ref{eq:3.1}) relates the charges at 
{\em the same magnetic field}.
It is also important to emphasize that Eq.~(\ref{eq:3.1}) is valid for
{\em arbitrary strength} of the pumping potential.
It is not restricted by 
 the
bilinear response regime.
 Eq.~(23) of Ref.~\onlinecite{Zhou99}       
 \ $Q_{\hookrightarrow}(B) =- Q_{\hookleftarrow}(-B)$  
together with  Eq.~(\ref{eq:3.1})
yield
 
\[
  Q(B)=Q(-B),
\]
where 
the pumping is performed along the same contour in the parameter
space. We intend to prove that,
unlike for the two-terminal conductance\cite{Onsager,Bonsager},
{\em such symmetry is not valid}.

The exact (not-averaged) $S$ - matrix of the system changes with
reversal of the magnetic field, $B$, as\cite{Landau3}:
\begin{equation}
{\cal S}(E, B)=\left[{\cal S}(E,- B)\right]^T.
\label{eq:3.2}
\end{equation}
Symmetry relation for the two terminal conductance (\ref{eq:2.1})
follows directly from Eq.~(\ref{eq:3.2}) \cite{Bonsager}
\begin{eqnarray*}
&&G(-B) \propto 
{\mathrm Tr}\left\{
\hat{\tau}_l
\hat{\cal S}(-B)\hat{\tau}_r
\hat{\cal S}^\dagger(-B)\right\}=\\
&&=
{\mathrm Tr}\left\{
\hat{\tau}_l
\hat{\cal S}^T(B)\hat{\tau}_r
\hat{\cal S}^*(B)\right\} 
=
{\mathrm Tr}\left\{
\hat{\tau}_r
\hat{\cal S}(B)\hat{\tau}_l
\hat{\cal S}^\dagger(B)\right\}=\\
&&=
{\mathrm Tr}\left\{
\hat{\tau}_l
\hat{\cal S}(B)\hat{\tau}_r
\hat{\cal S}^\dagger(B)\right\}
\end{eqnarray*}
 We omitted factors independent of the magnetic field in the
intermediate steps and 
used the unitarity of the $S$ - matrix. Therefore,
the relation
\[
G(B) = G(-B)
\]
is exact.

Now let us turn to the pumped charge. 
Substituting Eq.~(\ref{eq:3.2}) into Eq.~(\ref{eq:2.3b}), we find
\begin{eqnarray}
\Pi \left(-B\right)&=&
{\mathrm Im}\ {\mathrm Tr}\left\{\hat{\tau}_{l}
\frac{\partial \hat{\cal S}^T(B
)}{\partial X_2}
\frac{\partial \hat{\cal S}^*
(B)}{\partial X_1}
\right\}\label{eq:3.3}\\
&=&
{\mathrm Im}\ {\mathrm Tr}\left\{\hat{\tau}_{l}
\frac{\partial \hat{\cal S}^\dagger (B
)}{\partial X_1}
\frac{\partial \hat{\cal S}
(B)}{\partial X_2}
\right\}
= \Pi \left(B\right) \nonumber\\
&+&
{\mathrm Im}\ 
{\mathrm Tr}\left\{
\hat{\tau}_{l}
\left[
\frac{\partial \hat{\cal S}^\dagger (B)}{\partial X_1};
\frac{\partial \hat{\cal S}(B)}{\partial X_2}
\right]
\right\}
\nonumber
\end{eqnarray}
From Eq.~(\ref{eq:3.3}) one obtains,
\begin{equation}
\Pi \left(B\right) - \Pi \left(-B\right)=
\frac{i}{2}\varepsilon^{ij} {\mathrm Tr}\left\{
\hat{\tau}_{l}
\left[
\frac{\partial \hat{\cal S}^\dagger (B)}{\partial X_i};
\frac{\partial \hat{\cal S}(B)}{\partial X_j}
\right]
\right\}.
\label{eq:3.4}
\end{equation}
Commutator in right-hand-side of Eq.~(\ref{eq:3.4}) vanishes only if
$S$-matrix is symmetric. This is not the case in the presence
of a finite magnetic
field. Therefore, there is no fundamental symmetry
guarding the relation $\Pi \left(B\right) = \Pi \left(-B\right)$ as it
was in the case for the two-terminal conductance, and 
the Eq.~(23) of Ref.~\onlinecite{Zhou99}
 {\em does not} hold.
One can argue that it is the choice of the direction of the contour
 in the parameter space that violates the T - invariance
and thus, $B\to -B$ symmetry.

The absence of the symmetry with respect to the reversal
of the magnetic field, suggests that the 
correlation function $\langle
Q(B_1)Q(B_2)\rangle$ depends on the difference $B_1-B_2$ only and
vanishes at $|B_1 -B_2| \to \infty$
(as it does in generic parametric statistics).
 Model calculation of the following
subsection confirms this expectation.

\subsection{Model calculation of the second moment}
\label{sec:3.2}
In order to include the magnetic field into our description we have to
lift the condition that matrix ${\cal H}_{mn}$ from
Eqs.~(\ref{eq:2.6})--(\ref{eq:2.7}) is symmetric:
\[
{\cal H}_{mn} \to {\cal H}_{mn} + a {\cal H}_{mn}^a,
\]
where ${\cal H}_{mn}^a$ is the random realization of antisymmetric
$M\times M$ matrix and $a$ is the parameter proportional to the
magnetic field.  The resulting correlation function of two
Hamiltonians at different values of the magnetic field $B_{1,2}$ can be
conveniently presented in a form similar to Eq.~(\ref{eq:2.7}) as
\begin{eqnarray} 
\label{eq:3.5}
&&\langle{\cal H}_{nm}(B_1){\cal H}^*_{n'm'}(B_2)\rangle
=\left(\frac{\delta_1}{2\pi}\right)^2\times\\
&&\times\left[\delta_{nn'}\delta_{mm'}
\left(4M - N_h^D)
+\delta_{mn'}\delta_{nm'}
\left(4M - N_h^C\right)
\right]\nonumber
\right.
\end{eqnarray}
Quantities $N_h^{D,C}$ characterize the effect of the magnetic field
on the wave-functions of the closed dot and can be estimated as
\begin{equation}
N_h^{D} = g_{dot}
\left(\frac{\Phi_1 - \Phi_2}{\Phi_0}\right)^2,\quad
N_h^{C} = g_{dot}
\left(\frac{\Phi_1 + \Phi_2}{\Phi_0}\right)^2
\label{eq:3.6}
\end{equation}
where  $g_{dot}\gg 1$ is the dimensionless conductance of the closed
dot, $\Phi_{1(2)}$ is the magnetic flux through the dot which corresponds
to the
magnetic field $B_{1(2)}$, and $\Phi_0 = hc/e$ is the flux quantum.

For the Green functions in Eq.~(\ref{eq:2.18}) taken at
different magnetic fields  $B_1,B_2$,
 the diffuson, 
Eq.~(\ref{eq:2.19}), is  modified as
\begin{equation}
\label{eq:3.7}
{\cal D}
=\left(\frac{2\pi}{\delta_1}\right)^2\frac{1}
{
-i\epsilon + N_{ch} +N_h^{D}+ 
i\mbox{\boldmath $Z$}\cdot \mbox{\boldmath $X$}
+\mbox{\boldmath $X$}\hat{C}_0\mbox{\boldmath $X$}
}
\end{equation}

Diagrammatic representation, Fig.~\ref{Fig4}, and the expression for
Hikami box (\ref{eq:2.25}) remain intact.
Instead of Eq.~(\ref{eq:2.28}) we obtain
\begin{eqnarray}
&&\langle Q_A(B_1)Q_B(B_2)\rangle
=- \frac{e^2 N_lN_r}{N_{ch}^2}\int_adx_1
dx_2
\int_{b}dy_1dy_2 \nonumber\\
&&\times 
\left(\frac{\partial}{\partial \mbox{\boldmath $x$}_-}\right)^2
{\cal K}
\left(
\frac{2\pi T}{N_{ch}\delta_1},
\mbox{\boldmath $z$}\cdot\mbox{\boldmath $x$}_-,
\mbox{\boldmath $x$}_-^2 +\frac{N_h^D}{N_{ch}} 
\right),\nonumber\\
&&\mbox{\boldmath $x$}_-=
\mbox{\boldmath $x$}-\mbox{\boldmath $y$},
\label{eq:3.8}
\end{eqnarray}
Here the variables $\mbox{\boldmath $x$}, \mbox{\boldmath $y$},
\mbox{\boldmath $z$}.$ are determined by Eq.
(\ref{eq:2.27})  and the 
function ${\cal K}$ is given by Eq.~(\ref{eq:2.29}).

Equation (\ref{eq:3.8}) is the main result of this section. It
describes the sensitivity of the 
pumped charge to the magnetic field. One
immediately realizes that the correlation function depends only on the
difference of the magnetic fields in accord with the
discussion in the previous subsection.  Moreover the variance
of the charge does not depend on the magnetic field (for bilinear
response this result was obtained by Brouwer\cite{Brouwer99}).  
In terms of the diagrams, the absence of the symmetry with respect to the
magnetic field reversal is revealed in the fact that the Cooperon
does not contribute to the second moment even in the orthogonal case, $(B=0)$.

We conclude this section 
by the discussion of the asymptotics of
the correlation function $\langle Q_A(B_1) Q_A(B_2)\rangle$
at finite magnetic field. We
start with the weak pumping 
at low temperatures  $T \ll \delta_1N_{ch}$.
Instead of
Eq.~(\ref{eq:2.31}) one obtains
\begin{eqnarray}
\label{eq:3.9}
&&\langle Q_A(B_1)Q_A(B_2)\rangle = \\
&&=
 \frac{16e^2 N_lN_rS_a^2}{\pi^2N_{ch}^2}
\left[
\frac{2N_{ch}^3}{\left(N_{ch}+N_h^D\right)^3}+
\frac{N_{ch}^4\mbox{\boldmath $z$}^2    }
{\left(N_{ch}+N_h^D\right)^4}
\right],\nonumber 
\end{eqnarray}
where $N_h^D$  is related to the magnetic fields
by Eq.~(\ref{eq:3.6}).
If the dimensionless pumping potential is large ($
{\mathbf x}^2\gg \left(N_h^D/N_{ch}+1\right)$)
and the average level velocity is small, $\mbox{\boldmath $Z$}=0$,
we obtain instead of Eq.~(\ref{eq:2.36}):
\begin{equation}
\langle Q_A^2\rangle
= e^2 {\ell}_a \left(\frac{ 4N_lN_r}{\pi N_{ch}^2}\right)
\left(\frac{N_{ch}}{N_{ch}+N_h^D}\right)^{3/2}
.
\label{eq:3.10}
\end{equation}
At high magnetic field, $\langle Q_A(B) Q_A(-B)\rangle$ rapidly
decreases as $B^{-6}$, and one can use Eq.~(\ref{eq:3.9})  for the
bilinear response. 

Finally, we discuss the variance of the pumped charge 
in the magnetic field in the limit of
high temperatures 
($T\gg {\mathrm max}
(\frac{\delta_1(N_{ch}+N_h^D)}{2\pi},{\mathbf X}^2) $).
For the weak pumping, assuming the average level
velocity is small, we find
\begin{equation}
\label{eq:3.11}
\langle Q_A^2\rangle = 
 \left(\frac{4e^2 N_lN_rS_a^2}{3\pi N_{ch}^2}\right)
\left(\frac{N_{ch}\delta_1}{\pi T}\right)
\left(\frac{N_{ch}}
{N_{ch}+N_h^D}\right)^2
. 
\end{equation}
In the strong pumping regime
the high temperature
asymptotic behavior instead of Eq.~(\ref{eq:2.36}) is given by:
\begin{equation}
\langle Q_A^2\rangle
= e^2 {\ell}_a
\left( \frac{ 2N_lN_r}{ 3 N_{ch}^2}\right)
\left(\frac{N_{ch}\delta_1}{\pi T}\right)
\left(\frac{N_{ch}}{N_{ch}+N_h^D}\right)^{1/2}
.
\label{eq:3.12}
\end{equation}

\section{Discussion and conclusions}
\label{sec:4}
 Our main results include 
dependence on the pumping strength, temperature and magnetic field. 

{\em Dependence of the pumping strength} --- At the small pumping
potential, we essentially reproduced the results for bilinear
response\cite{Brouwer99,Zhou99}, that 
$\langle Q^2 \rangle \propto S_{A}^2$ with $S_{A}$ area being the {\em
area}
enclosed by the contour in the parametric space, see Eq.~(\ref{eq:2.31}).
This bilinear response regime, however, is valid only 
as long as the pumped
charge is smaller than unity. The regime of strong pumping is analyzed
for the first time in the present paper, see Eqs.~(\ref{eq:2.34}) ---
(\ref{eq:2.38}). This regime is hallmarked by the dependence
$\langle Q^2 \rangle \propto \ell_{A}$, with $\ell_{A}$  being the {\em
length} of the contour, which is substantially slower than naive expectation
$\langle Q^2 \rangle \propto S_{A}$. This  slow dependence was
already observed in Ref.~\onlinecite{Marcus99}. We think that
our conclusion about independence of the pumped charge variance on the
shape of the contour deserves a careful check\cite{Private}. 

{\em Temperature dependence} --- Our results for the high temperature
regime $T \gtrsim N_{ch}\delta_1$ indicate that the variance of the
charge $\langle Q^2\rangle$ is inversly proportional to the temperature
$\langle Q^2\rangle\propto 1/T$. Experiment\cite{Marcus99}
demonstrates $\langle Q^2\rangle \propto 1/T^2$ in the
high-temperature regime. This discrepancy was attributed to the
presence of the temperature dependent dephasing, ignored in our
treatment. In the simplistic models\cite{simpletons,dephexp}, the
dephasing is described by adding an extra factor $N_\varphi$ into the
mass of diffuson and Cooperon (\ref{eq:3.7}). If such a questionable
procedure is adopted, the effect
of dephasing would be described by replacement $N^D_h \to N^D_h+
N_\varphi$ in formulas of Sec.~\ref{sec:3.2}. We can see from
Eqs.~(\ref{eq:3.11}) --- (\ref{eq:3.12}) that the same $N_\varphi$ produces
different temperature dependences for the different regimes. Use of
experimentally\cite{dephexp} known dependence $N_\varphi \propto T$,
would produce the results $\langle Q^2\rangle \propto 1/T^3$ and
$\langle Q^2\rangle \propto 1/T^{3/2}$ for weak and strong pumping
respectively. We believe that the available experimental information
is not sufficient yet for making detailed comparison with our theory.

{\em Effect of the magnetic field} --- We have demonstrated in
Sec.~\ref{sec:3.1} that there is no fundamental reason
for the pumped current to be symmetric 
 with respect to the magnetic field
reversal, in a sharp contrast with the dependence of conductance on the
magnetic field. The corresponding correlation functions were calculated
in Sec.~\ref{sec:3.2}.  It is demonstrated there that $\langle
Q(B)Q(-B)\rangle \propto B^{-6}$ at large $B$. These conclusions
contradict to Ref.~\onlinecite{Marcus99} where the symmetry with
respect to magnetic field reversal was reported. We can not explain
this symmetry within the framework of our theory.

\acknowledgements
We are grateful to B. Spivak, C. Marcus, and F. Zhou 
for interesting discussions.  
I.A. was supported by
 A.P.~Sloan research foundation and Packard research foundation.
The work at Princeton University was supported by ARO
MURI DAAG 55-98-1-0270.

\appendix

\section{}

\label{app1}

We demonstrate the equivalence  between the equations
~(\ref{eq:2.3}) and the approach of Ref.~\onlinecite{Zhou99}.
The demonstration 
 will be based on equation of motion for
the Green functions (\ref{eq:2.12}) related to the $S$ - matrices by
Eq.~(\ref{eq:2.11}). 
First, we substitute Eq.~(\ref{eq:2.11}) into Eq.~(\ref{eq:2.13}), and
obtain
\begin{eqnarray}
\Pi &&\left(E,\mbox{\boldmath $X$}\right)
= 2i\pi^2\nu^2\epsilon^{ij} 
\frac{\partial^2}{\partial X_i^A \partial X_j^R} 
\label{eq:2.39}\\
&&\times {\mathrm Tr}
\left\{\hat{W}\hat{\tau}\hat{W}^{\dagger}
\hat{G}^R\left(E, \mbox{\boldmath $X$}^R\right)
\hat{W}\hat{W}^{\dagger}
\hat{G}^A\left(E, \mbox{\boldmath $X$}^A\right)
\right\}. \nonumber
\end{eqnarray}
We will show now that matrix
$\hat{G}^R\hat{W}\hat{W}^{\dagger}\hat{G}^A$ can be simplified
significantly using the equations for the Green functions
(\ref{eq:2.12}). We pre-multiply Eq.~(\ref{eq:2.12}) for $\hat{G}^A$
by $\hat{G}^R$, we transpose Eq.~(\ref{eq:2.12})  for
$\hat{G}^R$ and post-multiply it by $\hat{G}^A$. Subtracting the
results, we find
\begin{eqnarray}
2i&&\pi\nu
\hat{G}^R\left(E, \mbox{\boldmath $X$}^R\right)
\hat{W}\hat{W}^{\dagger}
\hat{G}^A\left(E, \mbox{\boldmath $X$}^A\right)
\label{eq:2.40}\\
&&=\hat{G}^A\left(E, \mbox{\boldmath $X$}^A\right)-
\hat{G}^R\left(E, \mbox{\boldmath $X$}^R\right)
\nonumber\\
&&+ \hat{G}^R\left(E, \mbox{\boldmath $X$}^R\right)
\left[
{\hat H}\left(\mbox{\boldmath $X$}^R\right) - 
{\hat H}\left(\mbox{\boldmath $X$}^A\right)\right]
\hat{G}^A\left(E, \mbox{\boldmath $X$}^A\right).
\nonumber
\end{eqnarray}
Substituting Eq.~(\ref{eq:2.40}) into (\ref{eq:2.39}), we obtain
\begin{eqnarray}
&&\Pi\left(E,\mbox{\boldmath $X$}\right)
=  
\label{eq:2.41}\\
&&\epsilon^{ij}\frac{\partial}{\partial X_i}  {\mathrm Tr}
\left\{\pi\nu \hat{W}\hat{\tau}\hat{W}^{\dagger}
\hat{G}^R\left(E, \mbox{\boldmath $X$}\right)
\frac{\partial \hat{H} \left(\mbox{\boldmath $X$}\right)}
{\partial X_j}
\hat{G}^A\left(E, \mbox{\boldmath $X$}\right)
\right\}. \nonumber
\end{eqnarray}
If we recall that $\pi\nu\hat{W}\hat{\tau}\hat{W}^{\dagger}$ is an
operator of the current from the dot through the left contact, 
we obtain Eq.~(10) of Ref.~\onlinecite{Zhou99}, which proves that the
physical mechanisms considered in Ref.~\onlinecite{Brouwer99} and
\onlinecite{Zhou99} are identical.

\end{multicols}
\end{document}